# A waveguide kinetics framework for electrochemical polarization

Bishuang Chen[a] and Huayang Cai[b,c,d,*]

Author affiliations: [a]School of Marine Sciences, Sun Yat-Sen University, Zhuhai 519080, China; [b]Institute of Estuarine and Coastal Research, School of Ocean Engineering and Technology, Sun Yat-Sen University / Southern Marine Science and Engineering Guangdong Laboratory (Zhuhai), Zhuhai, Guangdong 519082, China; [c]State and Local Joint Engineering Laboratory of Estuarine Hydraulic Technology / Guangdong Provincial Engineering Research Center of Coasts, Islands and Reefs / Guangdong Provincial Key Laboratory of Marine Resources and Coastal Engineering / Guangdong Provincial Key Laboratory of Information Technology for Deep Water Acoustics / Key Laboratory of Comprehensive Observation of Polar Environment (Sun Yat-Sen University), Ministry of Education, Zhuhai, Guangdong 519082, China; [d]Zhuhai Research Center, Hanjiang National Laboratory, Zhuhai, Guangdong 519082, China.

[*]Corresponding author: Huayang Cai.

**Email:** caihy7@mail.sysu.edu.cn

**Abstract**

Hydrogen electrocatalysis (HER/HOR) exhibits an anomalous, non-Nernstian pH dependence that has motivated competing mechanistic narratives yet still lacks a unified, transferable, and quantitatively predictive description across conditions. Here, we introduce a theory-neutral waveguide kinetics framework that reinterprets the polarization curve as a power-flow-like response, enabling a compact modal representation of interfacial kinetics. Without presuming any specific mechanism, the model quantitatively fits four representative polarization datasets historically explained by divergent theories. From each fit, we extract interpretable diagnostics, including a reflection amplitude and a useful-output density, that provide transferable metrics for interfacial efficiency. The framework thus establishes a computational experiment platform for mechanistic triangulation, operating-regime diagnosis, and the rational design of hydrogen electrocatalysts across the pH spectrum.

**Introduction**

Hydrogen electrocatalysis, encompassing the hydrogen evolution and oxidation reactions (HER/HOR), is foundational to water electrolyzers and fuel cells (1). A persistent, mechanistically opaque challenge is its anomalous non-Nernstian pH dependence, where kinetics in alkaline media are orders of magnitude slower than in acid (2). This pH effect has spurred competing microscopic narratives, including hydrogen-binding energy, bifunctional, interfacial water structure, and cation-coordination theories (3). While each finds experimental support, the field remains divided, lacking a consensus on the dominant descriptor. Fundamentally, these qualitative or semi-quantitative theories can rationalize data retrospectively but struggle to provide a unified, predictive quantitative framework for polarization behavior across diverse conditions (3,4).

This predictive gap has motivated multiscale computational models that integrate microkinetics and mass transport (5). However, such models typically embed a single mechanistic hypothesis, serving as validation tools for specific theories rather than neutral platforms for comparative evaluation (6-9). Their implementations vary, making fair cross-condition comparison and transferability difficult (10).

Here, we develop a theory-neutral waveguide kinetics framework to meet this need. It conceptually treats the polarization curve as a power-flow response, enabling a compact, analytic representation of kinetics constrained by passivity. This approach naturally captures essential features (exponential regimes, saturation, and asymmetry), while outputting mechanism-agnostic diagnostics like modal weights and a reflection amplitude. By successfully fitting diverse polarization datasets historically explained by competing theories, the framework establishes a transferable computational experiment platform. This platform allows for the quantitative comparison of systems, the diagnosis of dominant operating regimes, and ultimately provides a unified lens to guide catalyst and electrolyte design across the pH spectrum.

**Results**

**A waveguide kinetics framework for polarization analysis.** To quantitatively analyze polarization behaviors beyond specific mechanistic assumptions, we developed a theory-neutral waveguide kinetics framework based on a waveguide mass-energy relation we proposed previously (11). This framework reconceptualizes the steady-state polarization curve $j(\eta)$ as a power-flow-like response of the electrochemical interface to the driving overpotential $\eta$.

The core of the model is a dimensionless, passivity-constrained representation. The applied overpotential is first mapped to a dimensionless driving coordinate $\tilde{\eta}$, defined as $\tilde{\eta} = f\,\eta_{\text{eff}}, f = \frac{\mathcal{F}}{RT}$, where $\eta_{\text{eff}}$ is the effective interfacial overpotential (and reduces to the applied $\eta$ when no additional corrections are used), $\mathcal{F}$ is the Faraday constant (96485 C/mol), $R$ is the universal gas constant (8.314 J/mol·K), and $T$ is the temperature. In datasets where an Ohmic correction ($R_{\text{ohm}}$) is applied, $\eta_{\text{eff}} = \eta - jR_{\text{ohm}}$.

The kinetics are then described by a superposition of fundamental modes. Each mode $p$ is defined by a pair of non-negative forward and backward fluxes, denoted by the subscripts 'ox' (oxidation-directed/forward flux) and 'red'(reduction-directed/backward flux):

$$J_{\text{ox},p}(\tilde{\eta}) = \alpha_p \exp(\xi_p \lambda_p \tilde{\eta}), J_{\text{red},p}(\tilde{\eta}) = \alpha_p \exp(-(1-\xi_p)\lambda_p \tilde{\eta}), \tag{1}$$

where $\alpha_p$ is the mode strength, $\xi_p \in (0,1)$ controls intrinsic asymmetry, and $\lambda_p > 0$ is a growth-rate parameter. The net current-like response for a mode is derived from the flux imbalance: $\tilde{j}_p = (J_{\text{ox},p} - J_{\text{red},p})/\sqrt{1 + (J_{\text{ox},p} - J_{\text{red},p})^2}$.

Realistic interfaces often involve concurrent processes. We therefore model the total response as a signed two-mode superposition:

$$\tilde{j}_{\text{tot}}(\tilde{\eta}) = \tilde{j}_A(\tilde{\eta}) + \sigma_B \tilde{j}_B(\tilde{\eta}), \tag{2}$$

where $\sigma_B = +1$ for cathodic reduction-dominant systems (e.g., HER), and $\sigma_B = -1$ for anodic oxidation-dominant systems (e.g., HOR), where the second mode represents a compensating, current-suppressing process (Fig. 1A, B).

A key diagnostic output of the framework is the generalized reflection amplitude $\rho$, computed from the flux imbalance:

$$\rho(\tilde{\eta}) = \sqrt{\frac{\min(J_{\text{ox}}, J_{\text{red}})}{\max(J_{\text{ox}}, J_{\text{red}})}} \in [0,1]. \quad (3)$$

This metric quantifies the bidirectional feedback strength at the interface. A value of $\rho \approx 1$ indicates near-balanced forward and backward fluxes (strong feedback, akin to a standing wave), while $\rho \to 0$ signifies highly unidirectional, transmission-dominant operation (Fig. 1C, D).

To bridge this state diagnostic to practical operation, we define a useful-output density metric:

$$\Pi_{\text{dens}}(\tilde{\eta}) = \frac{|S_{\text{tot}}(\tilde{\eta})| [1 - \rho^2(\tilde{\eta})]}{|\tilde{\eta}| + \eta_0}, \quad (4)$$

where $S_{\text{tot}} = J_{\text{ox,tot}} - J_{\text{red,tot}}$ is the net power-flow flux, and $\eta_0$ is a small regularization constant. $\Pi_{\text{dens}}$ represents the effective power transfer per unit driving cost, accounting for both output magnitude and internal feedback losses. The maximum of $\Pi_{\text{dens}}$ on the dominant branch identifies a density-optimal operating point $\tilde{\eta}^*$, which defines the most efficient sweet spot for the system (Fig. 1E, F).

**Quantitative fitting and cross-condition diagnosis of experimental data.** We applied the framework to four distinct polarization datasets; each historically associated with a different mechanistic theory for the pH effect in hydrogen electrocatalysis. The model, without presuming any specific microscopic pathway, successfully fits all datasets quantitatively (Fig. 2A, C, E, G): (i) the broad-pH HER on Au(111), capturing the kinetic-to-mass-transport transition in dilute acid and convergent alkaline kinetics (12); (ii) the enhanced activity of bifunctional catalysts designed for water dissociation (13); (iii) the activity modulation by organic additives that alter the interfacial hydrogen-bond network (14); and (iv) the sluggish alkaline HOR on Pt-based catalysts and its improvement upon alloying, where the model automatically identifies the signature of anodic suppression ($\sigma_B = -1$) (15).

The quantitative fits are robust across all systems, with root-mean-square errors consistently below 5 % of the respective current-density ranges. The fitted parameters span physically interpretable intervals: mode strengths ($\alpha_p$) vary from ~$10^{-7}$ to ~$10^2$, covering orders-of-magnitude activity differences; intrinsic asymmetry coefficients ($\xi_p$) lie between 0.02 and 0.95, reflecting diverse balances between forward and backward fluxes; and growth-rate parameters ($\lambda_p$) range from 0.1 to 9, encoding the kinetic sensitivity to overpotential. From these fits we extract the density-optimal operating point $\tilde{\eta}^*$ and its corresponding output density $\Pi_{\text{dens}}^*$ for each system. For HER, $\tilde{\eta}^*$ shifts positively with increasing pH (e.g., from −0.22 V at pH 1 to −0.32 V at pH 13), quantifying the additional driving cost required in alkaline media. The concomitant rise in $\Pi_{\text{dens}}^*$ (e.g., from 0.26 at pH 1 to 0.37 at pH 13) indicates that, despite the higher overpotential, the systems can operate with improved power-transfer efficiency under optimized conditions.

The true power of the framework lies in its output of transferable diagnostics. By analyzing the reflection amplitude $\rho(\tilde{\eta})$ and the useful-output density $\Pi_{\text{dens}}(\tilde{\eta})$ (Fig. 2B, D, F, H), we can compare systems on a unified scale. The peak value of $\Pi_{\text{dens}}$ quantifies the best attainable efficiency of a system, while the position of the optimal operating point $\tilde{\eta}^*$ (marked by circles) indicates the driving cost required to reach it. For example, the analysis clearly reveals how a high-performing bifunctional catalyst operates at a less negative $\tilde{\eta}^*$ with a higher $\Pi_{\text{dens}}$ peak compared to a baseline, or how the HOR in alkali is penalized by a more positive optimal point and a lower peak density.

**Discussion**

Our waveguide kinetics framework addresses a central dichotomy in electrocatalysis: the need for mechanistic understanding versus the lack of a theory-neutral language for quantitative, transferable comparison. By mapping polarization to a power-flow response, we move beyond the paradigm of using computational models to validate a single hypothesis. Instead, the framework functions as a computational experiment platform, capable of integrating kinetic descriptors from competing theories, such as hydrogen binding, water dissociation, or interfacial structure, into a unified, analytically tractable form.

The key outcome is the extraction of mechanism-agnostic state diagnostics, particularly the reflection amplitude $\rho$ and the useful-output density $\Pi_{\text{dens}}$. These metrics provide a common vocabulary to quantify and compare interfacial efficiency across the acidic-alkaline divide. For instance, the consistent leftward shift of the density-optimal operating point $\eta^*$ for HER in alkaline media, quantified by our fits, offers a unified descriptor for the pervasive pH effect that is decoupled from any specific microscopic narrative. Similarly, the framework naturally identifies anodic systems where net current is suppressed ($\sigma_B = -1$), offering a quantitative signature for processes like the sluggish alkaline HOR, without presupposing its origin.

This approach bridges a critical gap between interpretative theory and predictive design. The diagnostics it provides serve as a quantitative fingerprint for any catalyst/electrolyte system. This enables researchers to objectively rank performance, diagnose the dominant operational regime, and identify the overpotential for maximally efficient operation. While the current model focuses on steady-state kinetics, its core principle of diagnosing systems via their driven response is general. We envision its extension to dynamically evolving interfaces and other electrocatalytic reactions plagued by mechanistic multiplicity, paving the way toward a more unified, diagnosis-driven methodology for electrocatalyst development.

**Materials and Methods**

Detailed methods can be found in the *SI Appendix*.

**Data, Materials, and Software Availability.** The four-dataset used in this study can be accessed at (12-15), respectively. The MATLAB scripts used for reproducing all figures are openly released at https://github.com/Huayangcai/waveguide-kinetics-framework-for-electrochemical-polarization.

**Acknowledgments**

This research was supported by the Guangdong Basic and Applied Basic Research Foundation (Grant No. 2023B1515040028), the National Natural Science Foundation of China (Grant No. 52279080, 42376097), and the Guangdong Provincial Department of Science and Technology (Grant No. 2019ZT08G090).

**Figures**

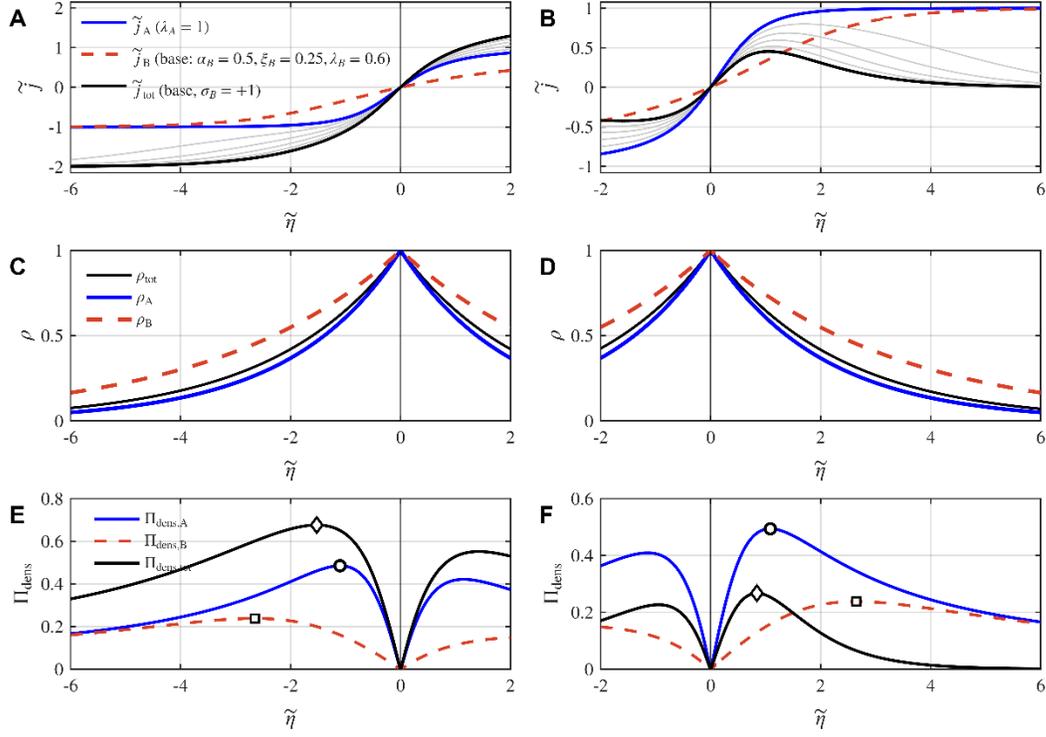

**Figure 1. Two-mode waveguide kinetics framework.**

(A, B) Net dimensionless polarization response $\tilde{j}_{\text{tot}}(\tilde{\eta})$ under signed superposition of a primary mode (A, blue) and a secondary mode (B, red dashed). Panels show cathodic ($\tilde{j}_A + \tilde{j}_B$) and anodic ($\tilde{j}_A - \tilde{j}_B$) cases. Grey traces illustrate modulation by the secondary mode's strength ($\alpha_B$);

(C, D) Corresponding generalized reflection amplitude $\rho(\tilde{\eta})$, a diagnostic of flux imbalance and interfacial feedback, for individual modes and the combined system;

(E, F) Derived useful-output density $\Pi_{\text{dens}}(\tilde{\eta})$, identifying the efficiency-optimal operating point $\tilde{\eta}^*$ (markers). Fixed parameters: $\alpha_A = 1$, $\lambda_A = 1$, $\lambda_B = 0.6$; see legend for $\xi$ and $\alpha_B$ values.

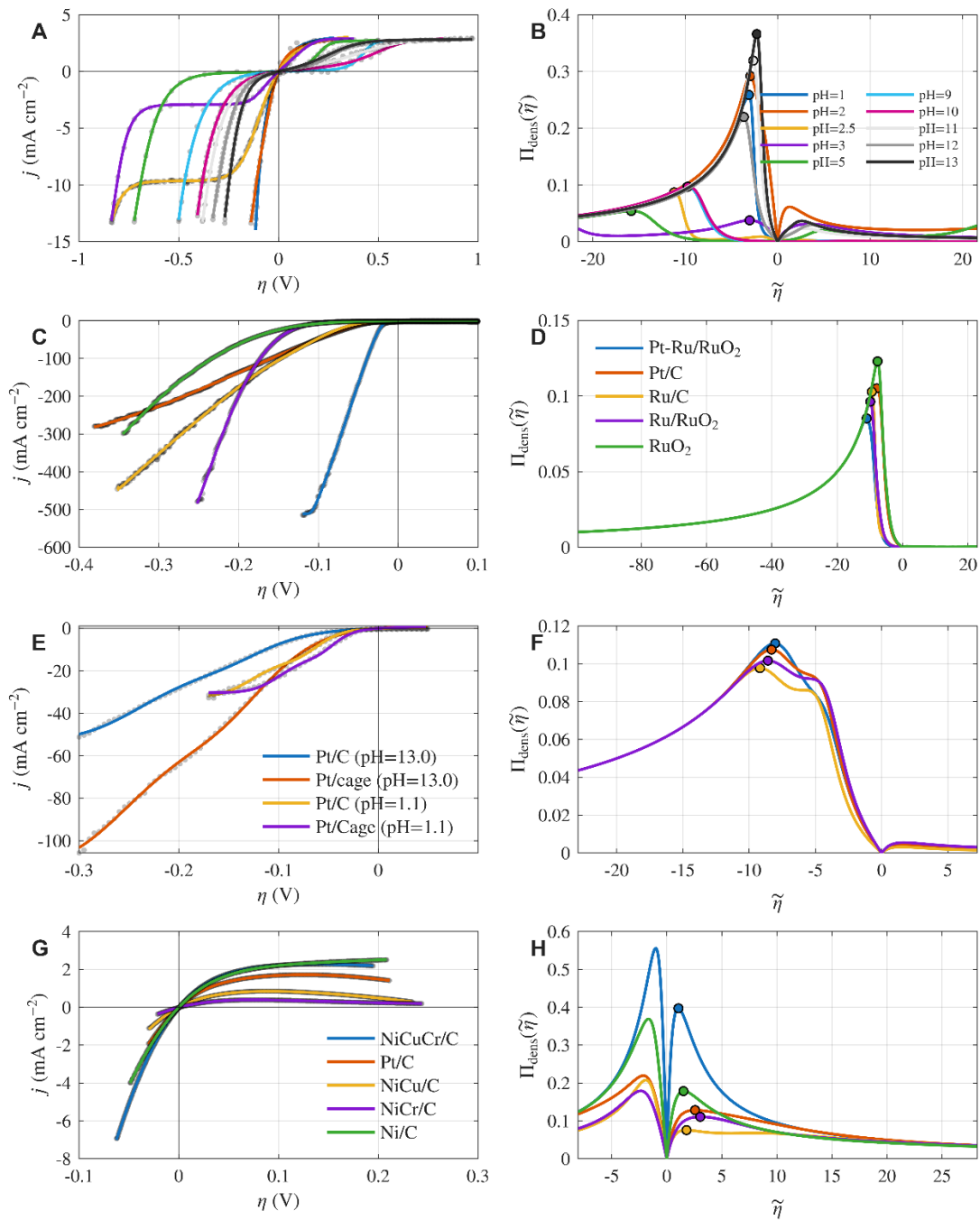

**Figure 2.** Waveguide-kinetics diagnostics and density-optimal operation extracted from experimental polarization datasets.

(A, C, E, G) Experimental polarization curves (grey markers) with two-mode waveguide-kinetics fits (colored lines) across four systems historically associated with different pH-effect theories;

(B, D, F, H) Corresponding useful-output density $\Pi_{\text{dens}}(\tilde{\eta})$ profiles, derived from the model's flux-based representation (see *SI* Appendix for definitions). The generalized reflection amplitude quantifies interfacial feedback, and $\Pi_{\text{dens}}$ integrates output magnitude and feedback loss. Filled circles denote the density-optimal operating points $\tilde{\eta}^*$, highlighting shifts in peak efficiency and required driving across pH and catalysts.

**Supporting Information for**

A waveguide kinetics framework for electrochemical polarization

Bishuang Chen[a] and Huayang Cai[b,c,d,*]

Corresponding author: Huayang Cai.

Email: caihy7@mail.sysu.edu.cn

**Supporting Information Text**

**Extended Methods**

**Waveguide-based polarization kinetics with two-mode superposition**

**Dimensionless driving coordinate and single-mode flux-based kinetics.** We work in the dimensionless overpotential $\tilde{\eta}$ defined by

$$\tilde{\eta} = f\, \eta_{\text{eff}}, f = \frac{\mathcal{F}}{RT}, \tag{S1}$$

where $\eta_{\text{eff}}$ denotes the effective interfacial overpotential (and reduces to the applied $\eta$ when no additional corrections are used), $\mathcal{F}$ is the Faraday constant (96485 C/mol), $R$ is the universal gas constant (8.314 J/mol·K), and $T$ is the temperature. When needed, the physical overpotential is recovered as $\eta_{\text{eff}} = \tilde{\eta}/f$.

For each mode $p \in \{A, B\}$ we define a non-negative forward/backward flux pair and its associated flux-based power-flow proxy,

$$J_{\text{ox},p}(\tilde{\eta}) = \alpha_p \exp(\xi_p \lambda_p \tilde{\eta}), J_{\text{red},p}(\tilde{\eta}) = \alpha_p \exp[-(1-\xi_p)\lambda_p \tilde{\eta}],$$
$$U_p = J_{\text{ox},p} + J_{\text{red},p}, S_p = J_{\text{ox},p} - J_{\text{red},p}, \tilde{\jmath}_p(\tilde{\eta}) = \frac{S_p}{\sqrt{1+S_p^2}}. \tag{S2}$$

Here $\alpha_p$ sets the mode strength, $\xi_p \in (0,1)$ controls intrinsic asymmetry between the two exponential branches, and $\lambda_p > 0$ is a growth-rate parameter that controls how rapidly the forward/backward imbalance develops with $\tilde{\eta}$. The bounded response $\tilde{\jmath}_p$ in Eq. (S2) is used as a convenient passive display variable for Row 1 of Fig. 1, while the feedback diagnostics below are evaluated directly from the non-negative flux pair $(J_{\text{ox},p}, J_{\text{red},p})$.

**Two-mode superposition for cathodic and anodic systems.** We model the net polarization response as a signed two-mode superposition,

$$\tilde{j}_{\text{tot}}(\tilde{\eta}) = \tilde{j}_A(\tilde{\eta}) + \sigma_B \tilde{j}_B(\tilde{\eta}),$$
$$\sigma_B = \begin{cases} +1, & \text{cathodic reduction-dominant system,} \\ -1, & \text{anodic oxidation-dominant system.} \end{cases} \quad (S3)$$

This signed superposition allows $|\tilde{j}_{\text{tot}}|$ to exceed unity because it represents the sum/difference of modal power-flow-like contributions rather than a globally normalized single-mode response. In the flux-based Scheme construction, the pair $(U, S)$ is treated as the physically meaningful (non-negative) throughput proxy and the signed power-flow flux, respectively, both inferred directly from the forward/backward directional fluxes. Accordingly, the combined-system power-flow channel is obtained by direct flux addition rather than by the bookkeeping sign $\sigma_B$:

$$U_{\text{tot}}(\tilde{\eta}) = U_A(\tilde{\eta}) + U_B(\tilde{\eta}), S_{\text{tot}}(\tilde{\eta}) = S_A(\tilde{\eta}) + S_B(\tilde{\eta}), \quad (S4)$$

equivalently $J_{\text{ox,tot}} = J_{\text{ox},A} + J_{\text{ox},B}$ and $J_{\text{red,tot}} = J_{\text{red},A} + J_{\text{red},B}$, so that $U_{\text{tot}} = J_{\text{ox,tot}} + J_{\text{red,tot}}$ and $S_{\text{tot}} = J_{\text{ox,tot}} - J_{\text{red,tot}}$. The sign parameter $\sigma_B = \pm 1$ in Eq. (S3) is therefore used only to classify cathodic versus anodic net-current superposition in $\tilde{j}_{\text{tot}}$.

**Generalized reflection amplitude from flux imbalance.** To quantify bidirectional feedback strength in a way directly tied to forward/backward imbalance, we define

$$\beta(\tilde{\eta}) = \frac{S(\tilde{\eta})}{U(\tilde{\eta})}, \rho(\tilde{\eta}) = \sqrt{\frac{1 - |\beta(\tilde{\eta})|}{1 + |\beta(\tilde{\eta})|}} = \sqrt{\frac{\min\{J_{\text{ox}}(\tilde{\eta}), J_{\text{red}}(\tilde{\eta})\}}{\max\{J_{\text{ox}}(\tilde{\eta}), J_{\text{red}}(\tilde{\eta})\}}} \in [0,1], \quad (S5)$$

evaluated for Mode A, Mode B, and the combined system by inserting $(U, S) = (U_A, S_A)$, $(U_B, S_B)$, and $(U_{\text{tot}}, S_{\text{tot}})$ from Eq. (S4). For the single-mode form in Eq. (S2) this construction yields the closed forms $\beta_p(\tilde{\eta}) = \tanh(\lambda_p \tilde{\eta}/2)$ and $\rho_p(\tilde{\eta}) = \exp(-|\lambda_p \tilde{\eta}|/2)$, showing that $\lambda_p$ directly controls the $\rho$-profile (hence the separation of $\rho_A$ and $\rho_B$ when $\lambda_A \neq \lambda_B$).

**Density-optimal operating point.** To connect the waveguide state diagnostics to an operational criterion, we evaluate a useful-output proxy and its density along $\tilde{\eta}$ using the reflection amplitude in Eq. (S5),

$$\Pi_{\text{use}}(\tilde{\eta}) = |S_{\text{tot}}(\tilde{\eta})| \, [1 - \rho^2(\tilde{\eta})], \quad \Pi_{\text{dens}}(\tilde{\eta}) = \frac{\Pi_{\text{use}}(\tilde{\eta})}{|\tilde{\eta}| + \eta_0}, \tag{S6}$$

where $\eta_0$ is a small regularization constant (here $\eta_0 = 0.01$ in dimensionless units) that prevents the density from being dominated by the $\tilde{\eta} \to 0$ singularity and stabilizes comparisons across conditions. We compute $\Pi_{\text{dens}}$ separately for Mode A, Mode B, and the total response (using their corresponding $\rho$ values) and locate the density-optimal points $\tilde{\eta}^*$ as the maximizers of $\Pi_{\text{dens}}$ restricted to the dominant branch (cathodic: $\tilde{\eta} < 0$; anodic: $\tilde{\eta} > 0$). In Fig.1A, B we fix Mode A with $\alpha_A = 1$ and $\lambda_A = 1$ (using $\xi_A = 0.35$ for the cathodic case and $\xi_A = 0.70$ for the anodic case) and prescribe Mode B with fixed asymmetry ($\xi_B = 0.25$ cathodic; $\xi_B = 0.75$ anodic) and a slower growth rate ($\lambda_B = 0.6$), while sweeping $\alpha_B$ to generate the grey family; the highlighted baseline uses the parameter values shown in the legend. Fig. 1C, D report $\rho_A(\tilde{\eta})$, $\rho_B(\tilde{\eta})$, and $\rho_{\text{tot}}(\tilde{\eta})$, revealing how differing $\lambda$ values reshape the feedback landscape under cathodic enhancement versus anodic suppression, and Figs. 1E, F show the corresponding $\Pi_{\text{dens}}$ curves and the operating-point shifts induced by the secondary mode.

**Definition of Waveguide-Kinetics Diagnostics**

For a given mode or their superposition, the total flux $U$ and net power-flow flux $S$ are defined from the forward ($J_{\text{ox}}$) and backward ($J_{\text{red}}$) components:

$$U = J_{\text{ox}} + J_{\text{red}}, \quad S = J_{\text{ox}} - J_{\text{red}}. \tag{S7}$$

The generalized reflection amplitude $\rho \in [0,1]$ is computed as:

$$\beta = \frac{S}{U}, \rho = \sqrt{\frac{1-|\beta|}{1+|\beta|}}, \tag{S8}$$

which quantifies the degree of flux imbalance (Fig. 1C, D). The useful-output density $\Pi_{\text{dens}}$, representing effective power transfer per unit driving cost (Fig. 1E, F and Fig. 2B, D, F, H), is given by:

$$\Pi_{\text{use}} = |S|(1-\rho^2), \Pi_{\text{dens}}(\tilde{\eta}) = \frac{\Pi_{\text{use}}(\tilde{\eta})}{|\tilde{\eta}|+\eta_0}, \tag{S9}$$

where $\eta_0 = 0.01$ is a small regularization constant. The density-optimal operating point $\tilde{\eta}^*$ is the maximizer of $\Pi_{\text{dens}}$ on the dominant branch of the polarization response.